\documentclass[12pt]{iopart}

\usepackage{graphicx}
\usepackage{iopams}
\usepackage{bm}
\usepackage[numbers]{natbib}
\usepackage{dsfont}
\usepackage{epsfig}
\usepackage{blindtext, rotating}
\usepackage{mathtools}
\usepackage{dsfont}
\usepackage{subcaption}
\usepackage{physics}
\usepackage{ragged2e}
\usepackage{siunitx}
\usepackage{comment}
\usepackage{lipsum}
\usepackage{hyperref} 
\hypersetup{breaklinks=true, colorlinks=true, citecolor=blue, linkcolor=cyan, urlcolor=blue,filecolor=blue}

\usepackage{xcolor} 

\newcommand\defequal{\stackrel{\mathclap{\tiny\mbox{def}}}{=}}

\DeclareCaptionJustification{justified}{\justifying}

\begin{document}

\title{The quantum optics of gravitational waves}

\author{Luca Abrah\~ao}
\ead{lucaabrahao@aluno.puc-rio.br}
\address{Department of Physics, Pontifical Catholic University of Rio de Janeiro, Rio de Janeiro 22451-900, Brazil}

\author{Francesco Coradeschi}
\ead{fr.coradeschi@gmail.com}
\address{Istituto del Consiglio Nazionale delle Ricerche, OVI, Italy}

\author{Antonia Micol Frassino}
\ead{antoniam.frassino@icc.ub.edu}
\address{Departament de F{\'\i}sica Qu\`antica i Astrof\'{\i}sica, Institut de Ci\`encies del Cosmos,
  Universitat de Barcelona, Mart\'{\i} i Franqu\`es 1, E-08028 Barcelona, Spain}

\author{Thiago Guerreiro}
\ead{barbosa@puc-rio.br}
\address{Department of Physics, Pontifical Catholic University of Rio de Janeiro, Rio de Janeiro 22451-900, Brazil}

\author{Jennifer Rittenhouse West}
\ead{jennifer@lbl.gov}
\address{Lawrence Berkeley National Laboratory, Berkeley, CA 94720, USA}

\author{Enrico Junior Schioppa}
\ead{enrico.junior.schioppa@chimicifisici.it}
\address{Via della libert\`a 36, 73025 Martano (LE), Italy}

\begin{abstract}
  %
  %
  By utilizing quantum optics techniques, we examine the characteristics of a quantum gravitational wave (GW) signature at interferometers. In particular, we study the problem by analyzing the equations of motion of a GW interacting with an idealized interferometer. Using this method, we reconstruct the classical GW signal from a representation of the quantum version of an \emph{almost classical} monochromatic wave (a single-mode coherent state), then we discuss the experimental signatures of some specific, more general quantum states. We calculate the observables that could be used at future interferometers to probe possible quantum states carried by the gravitational waves.

\end{abstract}

\maketitle

\section{Introduction}



Our understanding of gravity is simply and elegantly expressed through the theory of general relativity (GR). GR is a well-defined \emph{classical} (as in, non-quantum) theory and its study offers numerous fruitful avenues of research in various domains, ranging from cosmology to astrophysics to tabletop experiments. GR may of course also be viewed as the low energy limit of a \emph{quantum} field theory. While a full-blown quantum theory of gravity is as yet unavailable, it is possible to give an \emph{effective} quantum description of GR by treating the spacetime metric as a classical background, and perturbatively quantizing fluctuations around it   \cite{Donoghue:1994dn,Donoghue:2022eay}. The resulting effective theory is a quantum field theory of gravitons in curved spacetime, and its non-renormalizability is harmless at energies much below the Planck scale. Such an effective description may arise, for instance, as the intermediate energy limit of a more complete theory. The detection of gravitational waves -- long awaited theoretically and difficult to achieve experimentally -- has been a huge scientific achievement.  In the same vein, a natural next step is to search for experimental or astrophysical observations that can provide definitive proof of (or exclude) the existence of gravitons.

Recently, there has been an increasing interest in combining characteristic features of gravity with aspects of quantum optics.  Some interesting applications that would benefit from this union would be the detection of gravitational waves using cavities~\cite{Krisnanda2017, Bose2017, Marletto2017, Oniga2016, Carney2021b, Carney2021c, Streltsov2022, Westphal2021} or, alternatively, the analysis of possible quantum effects in the weak gravity regime~\cite{Biswas2022, Coradeschi2021, Mazzitelli:2023des}. While the former points in the direction of increasing sensitivity, the latter focuses more on the phenomenological advantages of considering such experimental setups. 

From the phenomenological point of view, a significant benefit of combining quantum optics and gravity is the possibility of experimentally probing the existence of the graviton as the mediator of the gravitational force. 
Indeed, while empirical evidence supports the existence of quantum particles, such as photons, gluons, W/Z bosons, and Higgs bosons, which are known to give rise to all known interactions according to the Standard Model, gravitons have not been directly observed (and have even been argued to be in principle non-observable~\cite{Dyson2013}), even though they are likely to be an ingredient of any quantum theory of gravity.

As already mentioned, the quantum field description of an interacting spin-2 particle, having GR as a classical limit, has been long well-known. Even though such a theory only works phenomenologically at low energy, a great deal of knowledge can be gained from just focusing on such a low-energy, weak-gravity (perturbative) quantum regime~\footnote{See also the detailed discussion regarding the definition of ``perturbative quantum gravity'' presented in~\cite{Cho:2021gvg}.}. From a practical point of view, this means that it makes sense to explore the possibility that weak gravitational phenomena, such as gravitational waves far from their sources, may have a quantum nature. 
Observations of gravitational waves by the LIGO-Virgo-KAGRA collaboration support classical general relativity, and quantum corrections, if any, are assumed to be negligible because graviton shot noise is suppressed by the large occupancy number of gravitons in a detectable wave \cite{Parikh2021a, Parikh2021b}. However, if quantum effects (negligible or not) are present, the argument for their smallness is only reliable if the gravitational wave is in a mostly classical state, such as a coherent state. For such states, it is unlikely that any quantum behaviour will be observed in the near future; however this conclusion does not apply, at least straightforwardly, to more general quantum states.

Recently, considerable effort has been devoted to understanding possible signatures at LIGO-like experiments of hypothetical quantum states of gravity that have no classical analog~\cite{Parikh2021a, Parikh2021b, Guerreiro2020, Guerreiro2021,Guerreiro:2023gdy} (see also~\cite{Bak:2022oyn, Cho:2021gvg, Kanno:2020usf}). 
These signatures can be fruitfully investigated by using an effective field theory description of gravity and using the formal tools of quantum optics. The picture that comes out gives a powerful perspective on some fundamental aspects of the possible phenomenology of quantum gravity.

We start by treating both gravity and the instrument one uses to detect it as quantum objects. We thus describe the dynamics of the system by a Hamiltonian operator of the form
\begin{equation}
  \label{eq:H0+HI}
  \hat{H} = \hat{H}_{0} + \hat{H}_{int}\,.
\end{equation}
where
\begin{equation}
  \hat{H}_{0} = \hat{H}_{g} + \hat{H}_{d}\,,
\end{equation}
gives the dynamics of pure gravity (subscript $g$) and of the detector (subscript $d$).
The details of $\hat{H}_{0}$ are not important for our discussion, beyond treating gravity as a quantum theory: the states we will consider are quantum superpositions of plane gravitational wave, that is, states of many gravitons of well-defined energy. The most interesting information lives instead in the interaction term $\hat{H}_{int}$, the form of which is very specific to the system we are studying. In other words, even though we are going to make use of the formal tools of quantum optics, which are generic no matter the specific form of $\hat{H}_{int}$, it is precisely the specific form of $\hat{H}_{int}$ that carries the differences between electromagnetism and gravity. This will allow us to draw conclusions that specifically pertain to the quantum weak regime of the gravitational field.

Let us now briefly outline the way it is possible to arrive at these conclusions with details to follow in the next sections.
As a first step, we formally prepare the gravitational part of our GW-detector system in some specific state $\ket{\Psi\left(t=0\right)}$ (we will consider several different set-ups for $\Psi$), then we choose a quantum observable $\hat{\mathcal{O}}$ and compute the outcome of (classical) measurements yielding expectation values of the form
\begin{equation}\label{eq:matrixElementGeneric}
  \mathcal{O}^{n}\left(t\right) = \mel{\Psi}{\hat{U}^{\dagger}\left(t\right) \hat{\mathcal{O}}^{n} \hat{U}\left(t\right)}{\Psi}\,,
\end{equation}
where $U$ is the time evolution operator
\begin{equation}
  \hat{U}\left(t\right) = e^{- \frac{i}{\hbar} \hat{H} t}\,,
\end{equation}
whose specific form for our GW-detector setup is known from previous results \cite{Guerreiro2020,Brandao2020}.
Calculating expectation values for different values of $n$, we can reconstruct the probability distribution for observable $\hat{\mathcal{O}}$ on state $\Psi$ in some detail (mean value, variance $\left[\Delta \mathcal{O}\left(t\right)\right]^2 = \mathcal{O}^2\left(t\right) - \left[\mathcal{O}\left(t\right)\right]^2$, and so on); we show that the moments of certain observables (in particular, the cavity's electric field $\mathcal{E}$) can depend strongly on the choice of $\Psi$, meaning that measurements of $\mathcal{O}^n$ can discriminate between different states $\Psi$ and -- potentially -- probe whether $\Psi$ possesses inherently quantum features that can't be mimicked by a purely classical GW.

The outline of the paper is the following.
In Section~\ref{sec:dynamics}, we make use of a few important previous results~\cite{Buonanno2003, Pang2018} to reduce the complicated starting forms of $\hat{H}_{d}$ and $\hat{H}_{int}$ to simple expressions.
Moving forward, we will focus on examining the changes in observables when the gravitational waveform $\ket{\Psi_{g}\left(t=0\right)}$ is prepared in specifically selected states in the remaining half of the Hilbert space.

In Section~\ref{sec:reconstruction}, 
we test our approach by applying the equations of motion derived in Sec.~\ref{sec:dynamics} to some example cases. First we examine the simplest allowed states for the gravitational field: the vacuum and the single-mode coherent state. Although the former does show some interesting quantum effects, they ultimately seem too small to be measured at present-day experiments. On the other hand, the latter serves to validate our program by providing the expected signal from a classical analog of a single mode coherent state: a classical plane monochromatic gravitational wave. Additional validation comes from Sec.~\ref{sec:decoherence}, where we show how the thermal gravitational background, when treated as 
an ensemble of mixed states, produces gravitational-induced decoherence~\cite{Guerreiro2021}. We thus recover known results from previous work~\cite{Blencowe2013}.

In Section~\ref{sec:observables}, we comment on the effect of GW fluctuations on electric fields, and how these can be used to obtain information on the GW state.

In Section~\ref{sec:squeezing}, we study hypothetical collective quantum states of gravitational waves with no classic analogue, and analyze their detection at an interferometer. We find that the gravitational \emph{squeezed vacuum} gives an effect on the \emph{noise} of the intereferometer~\cite{Parikh2021a, Guerreiro2020}, and that gravitational \emph{squeezed coherent} states affect its \emph{signal}, with some room left for such an effect being very significant~\cite{Guerreiro2021}. Needless to say, these conclusions remain hypothetical before we actually do see such a signal in an experiment. Nonetheless, and this is the topic of Section~\ref{sec:sources}, we give some heuristic arguments in favour of the existence of squeezed states in gravity by noting similarities and differences with the case of electromagnetism.

Finally, in Section~\ref{sec:conclusions} we give our conclusions.

\section{Dynamics of the system}\label{sec:dynamics}

We will be working within the framework of linearized gravity, employing a flat metric background to enable us to consider weak gravity far from sources.
With this choice, the quadratic part of the Einstein-Hilbert action (in vacuum) in the harmonic gauge $\partial_\mu h^{\mu\nu} = 0$ reduces to
\begin{equation}\label{eq:linearizedMinkowskiEHAction}
  S_{EH} = \frac{c^4}{32\pi G} \int d^4x \left( \frac{1}{2} \partial_\mu h_{\alpha\beta} \partial^\mu h^{\alpha\beta} - \frac{1}{4} \partial_\mu h \partial^\mu h \right)\,,
\end{equation}
where, as usual, the field $h^{\mu\nu}$ represents the small perturbations of the otherwise flat metric $\eta_{\mu \nu}$, and $h=\eta^{\mu\nu}h_{\mu\nu}$ is contracted by the flat metric tensor. We neglect higher order terms (that is, gravitational self-interactions) throughout, as their impact is negligible in GWs far from their source.


In the transverse traceless (TT) gauge, we expand the field into Fourier components as
\begin{equation}\label{eq:perturbationTT}
  h_{ij}^{TT} \left(t,\boldsymbol{x}\right) = \int \frac{d^3\boldsymbol{k}}{\sqrt{(2\pi)^3}} \epsilon_{ij}^\lambda \left(\boldsymbol{k}\right) h_\lambda\left(t,\boldsymbol{k}\right)e^{i\boldsymbol{k}\cdot\boldsymbol{x}}\,,
\end{equation}
where the $\epsilon_{ij}^\lambda \left(\boldsymbol{k}\right)$ are the tensors for the two polarization states $\lambda=+,\times$, satisfying the due conditions of orthonormality ($\epsilon_{ij}^\lambda\epsilon_{jk}^{\lambda'} = \delta_{ik}\delta^{\lambda\lambda'}$), transverseness ($\epsilon^\lambda_{ij} k^j  = 0$) and tracelessness ($\text{Tr}\left[\epsilon^\lambda_{ij}\right] = 0$). Notice that Greek indices have become Latin indices, as the time components of the field in the TT gauge are null ($h_{0\mu} = 0$).

With the field expressed in this form, we can execute canonical quantization by rewriting the field to operators. We promote the Fourier coefficients to annihilation and creation operators as follows
\begin{equation} \label{eq:operators_h}
  h_\lambda\left(t,\boldsymbol{k}\right) \quad \rightarrow \quad \hat{\mathfrak{b}}^\lambda_{\boldsymbol{k}}\,,
\end{equation}
\begin{equation}
  h^*_\lambda\left(t,\boldsymbol{k}\right) \quad \rightarrow \quad {\hat{\mathfrak{b}}^{\lambda \dagger}_{\boldsymbol{k}}}\,,
\end{equation}
which obey the standard commutation relations (from here on we work in units where $\hbar=1$),
\begin{equation}
  \left[\hat{\mathfrak{b}}^\lambda_{\boldsymbol{k}}, {\hat{\mathfrak{b}}^{\lambda' \dagger}_{\boldsymbol{k}'}} \right] = \delta_{\lambda\lambda'}\delta^{(3)}\left(\boldsymbol{k},\boldsymbol{k'}\right)\,. \label{eq:commutationRelations_b_1}
\end{equation}
The classical field now gets promoted to a quantum field operator and we can write explicitly
\begin{equation}
  \label{eq:hquantum}
  \hat{h}_{ij} \left(t,\boldsymbol{x}\right) = \int \frac{d^3\boldsymbol{k}}{\sqrt{(2\pi)^3}} \left( \sqrt{\frac{8\pi G}{k}} \epsilon_{ij}^\lambda \left(\boldsymbol{k}\right) \hat{\mathfrak{b}}^\lambda_{\boldsymbol{k}} e^{i\left(\boldsymbol{k}\cdot\boldsymbol{x}-\Omega_k t\right)} + h.c. \right)\,.
\end{equation}
Equation~\eqref{eq:hquantum} concludes the description of the setup we will be using for the gravitational part of our problem; we will mostly consider single-mode metric perturbations in the following, $\sim h_{\mu\nu}e^{i kx}$, corresponding to planar waves of well-defined frequency. This does not limit the scope of our calculations since we will be able to express any potential initial (quantum) state of gravity as a superposition of plane waves.

As far as the detector is concerned, we want to model a GW interferometer. Arvanitaki and Geraci~\cite{Arvanitaki2003} have shown that already a single-mode Fabry-P\'erot cavity is sensitive to gravitational waves. This can be achieved either by inserting a nanosphere in the setup~\cite{Arvanitaki2003}, or by letting one of the two mirrors be free of moving, as described by Buonanno and Chen in~\cite{Buonanno2003}, and by Pang and Chen in~\cite{Pang2018}. 
Let us narrow our focus to the second case of study. Pang and Chen have demonstrated, by making realistic assumptions (see~\cite{Pang2018}), that a complete model of a GW interferometer (including power recycling and signal recycling mirrors) can be mapped to a single Fabry-P\'erot cavity where one mirror is fixed and the other is free to move. This simplifies the complexity of the interferometer, and we can work with a single cavity of length $L_0$ as the only degree of freedom to describe our model detector.
%
%
When a GW of strain $h$ passes through such a cavity perpendicularly to its axis, its length changes in the following way:
\begin{equation}
  L_0 \quad \rightarrow \quad L_0\left(1+\frac{1}{2}h\right)\,.
\end{equation}
This can be seen as a ``gravitomechanical'' coupling between the GW and the detector, much like an optomechanical coupling between the electromagnetic field and a mechanical oscillator \cite{Aspelmeyer2014}.

Instead of working with the GW coupled to the detector's mirror, one can move to a perspective in which the GW couples directly to the cavity's electromagnetic field (the laser beam). When the cavity is stretched, its resonance frequency changes accordingly as
\begin{equation}
  \omega_0 = \frac{n\pi}{L_0} \quad \rightarrow \quad \omega = \omega_0 \, \frac{n\pi}{L_0\left(1+\frac{1}{2}h\right)}\,,
\end{equation}
which can be expanded as
\begin{equation}
  \omega = \omega_0 \left( 1-\frac{1}{2}h + \mathcal{O}\left(h^2\right) \right)\,.
\end{equation}
The induced frequency shift can be interpreted as producing an effective coupling between the GW and the electromagnetic field inside the cavity. It turns out, following~\cite{Guerreiro2020,Bassi2017}, that for a $+$ polarized GW propagating in the $z$ direction perpendicularly to the cavity axis ($x$ direction), and satisfying  $k_x L_0 \ll 1$,
such a coupling is represented by an interaction Hamiltonian of the form 
\begin{equation}\label{eq:HGWintContinuous}
  \hat{H}_\text{GW}^\text{int} = -\frac{\omega_0}{4} \hat{a}^\dagger\hat{a} \int \frac{d^3\boldsymbol{k}}{\sqrt{(2\pi)^3}} \left( \sqrt{\frac{8\pi G}{k}}  \hat{\mathfrak{b}}_{\boldsymbol{k}} + \text{h.c.} \right)\,.
\end{equation}
In this expression, $\hat{a}$ and $\hat{a}^\dagger$ are the annihilation and creation operators of the cavity field, which we take to be in a single mode state for simplicity, while the operators $\hat{\mathfrak{b}}$ are defined in~\eqref{eq:operators_h}. Notice that, having fixed the polarization of the GW, the $\lambda$ index has dropped.

Following a procedure which is standard in quantum optics~\cite{Scully1997}, we introduce a quantization volume $V$ to define a dimensionless quantity $\hat{b}_{\boldsymbol{k}} = \hat{\mathfrak{b}}_{\boldsymbol{k}}/ \sqrt{V}$,
and transform the continuous integral in Eq.~\eqref{eq:HGWintContinuous} into its discretized version~\cite{Guerreiro2020} 
\begin{equation}
  \hat{H}_\text{GW}^\text{int} = -\frac{\omega_0}{4} \hat{a}^\dagger\hat{a} \sum_{\boldsymbol{k}} \left( \sqrt{\frac{8\pi G}{Vk}}  \hat{b}_{\boldsymbol{k}} + \text{h.c.} \right)\,.
\end{equation}
\\
Let us now define, respectively, the single graviton strain $f_{\boldsymbol{k}}$, the opto-gravitational coupling constant $g_{\boldsymbol{k}}$, and the dimensionless coupling $q_{\boldsymbol{k}}$, in the following way:
\begin{equation}
  f_{\boldsymbol{k}} = \sqrt{\frac{8\pi G}{Vk}},\quad
  g_{\boldsymbol{k}} = \frac{\omega_0f_{\boldsymbol{k}}}{4},\quad
  q_{\boldsymbol{k}} = \frac{g_{\boldsymbol{k}}}{\Omega_k}\,. \label{eq:coupling}
\end{equation}
Here, $\boldsymbol{k}$ represents the GW frequency for the mode $\boldsymbol{k}$, where $|\boldsymbol{k}| = \Omega_k$. Since $q_{\boldsymbol{k}}$ is a small number by definition, we will treat it as a perturbative parameter.
%
With these definitions, the Hamiltonian for the complete system, including the GW, the cavity field, and their effective interaction, can be defined as
\begin{equation}\label{eq:H}
  \hat{H} = \hat{H}_0 + \hat{H}_\text{GW}^\text{int}\,,
\end{equation}
with the free Hamiltonian given by
\begin{equation}\label{eq:H0}
  \hat{H}_0 = \omega \hat{a}^\dagger\hat{a} + \sum_{\boldsymbol{k}} \Omega_{k} \hat{b}_{\boldsymbol{k}}^\dagger \hat{b}_{\boldsymbol{k}}\,,
\end{equation}
and the interaction Hamiltonian further reduced to
\begin{equation}\label{eq:InteractionTerm}
  \hat{H}_\text{GW}^\text{int} = - \hat{a}^\dagger\hat{a} \sum_{\boldsymbol{k}} q_{\boldsymbol{k}} \Omega_{\boldsymbol{k}} \left(   \hat{b}_{\boldsymbol{k}} + \hat{b}^\dagger_{\boldsymbol{k}} \right)\,.
\end{equation}
The derivation of the explicit form of the time evolution operator for the interaction term~\eqref{eq:InteractionTerm} is a lengthy but straightforward calculation, using an approach which is standard in quantum optics. As reported in~\cite{Guerreiro2020,Brandao2020}, we can express the result for a single mode $\boldsymbol{k}$ (omitting the index for the sake of readability) as
\begin{equation}\label{eq:evolutionOperatorFinal}
  \hat{U}\left(t\right) \ket{\Psi\left(t\right)} =
  e^{q\hat{a}^\dagger \hat{a} \left[ \eta\left(t\right)\hat{b} - \eta^*\left(t\right)\hat{b}^\dagger \right]}
  e^{i B\left(t\right)\left(\hat{a}^\dagger \hat{a}\right)^2} \ket{\Psi\left(t\right)}\,,
\end{equation}
where the time evolution is contained in the definition of
\begin{eqnarray}
  \eta\left(t\right) &=& 1-e^{-it}\,,\label{eq:definitionEta}\\
  \eta^*\left(t\right) &=& 1-e^{it}\,, \label{eq:definitionEta*}\\
  B\left(t\right) &=& q^2 \left(t-\sin\left(t\right)\right) \,,
  \label{eq:B}
\end{eqnarray}
and the time-evolving state that appears in~\eqref{eq:evolutionOperatorFinal} is defined as $\ket{\Psi\left(t\right)} = e^{-i \hat{b}^\dagger \hat{b}t} \ket{\Psi}$.

\section{GW state reconstruction}\label{sec:reconstruction}


When examining how a GW interacts with an optical cavity, the most suitable observable is the electric field operator that characterizes the cavity field's state
\begin{equation}
  \hat{\mathcal{E}} = \sqrt{\frac{\omega}{V_c}} \frac{\hat{a}+\hat{a}^{\dagger}}{\sqrt{2}}\,.
\end{equation}
This is indeed the physical quantity that one measures at an interferometer to produce a detectable signal. 
Note that here $V_c$ denotes the cavity mode volume. 

Now that we have defined the operator, we can proceed to calculate the matrix elements as outlined in Eq.~\eqref{eq:matrixElementGeneric}. Our focus will be on the mean value, specifically when $n=1$. The classical GW signal at an interferometer is sensed as a variation of the phase of the field quadrature:
\begin{equation}
  \mathcal{E} \quad \rightarrow \quad \mathcal{E} e^{i\phi}\,,
\end{equation}
where $\phi$ changes in time. For example, this variation of the phase produces typical chirp-like signatures observed for binary merger events. Any result we find that produces a departure of $\phi$ from its classical behavior, namely which has the form
\begin{equation}\label{eq:effectOnSignal}
  \mathcal{E} \quad \rightarrow \quad \mathcal{E} e^{i\left(\phi+\delta\phi\right)}\,,
\end{equation}
is interpreted as an effect on the \emph{signal}. Contrariwise, if we find a modification of the form
\begin{equation}\label{eq"effectOnNoise}
  \mathcal{E} \quad \rightarrow \quad \mathcal{E} + \epsilon\,,
\end{equation}
then we are witnessing an effect on the \emph{noise}.

To begin with, we observe that the expression for the time evolution operator~\eqref{eq:evolutionOperatorFinal} includes an exponential term in $q^2$ (the one defined as $B\left(t\right)$ in~\eqref{eq:B}). However, since the terms in $q$ is dominant, we can disregard this term for now (although we will revisit it in Sec.~\ref{sec:decoherence}).
Based on this assumption, we can use the time evolution of the electric field operator to obtain the following result
\begin{equation}\label{eq:electriFieldQuadratureOperatoreGWTraced}
  \mathcal{E}\left(t\right) = \sqrt{\frac{\omega}{V_c}}\left(\frac{\mel{\Psi\left(t\right)}{\hat{\mathcal{D}}\left[q\eta\left(t\right)\right]\hat{a}}{\Psi\left(t\right)} + h.c. }{\sqrt{2}}\right)\,,
\end{equation}
where we have defined the operator in parenthesis as
\begin{equation}\label{eq:displacementOperator}
  \hat{\mathcal{D}}\left[q\eta\left(t\right)\right] = e^{q\hat{a}^\dagger \hat{a} \left[ \eta\left(t\right)\hat{b} - \eta^*\left(t\right)\hat{b}^\dagger \right]}\,.
\end{equation}
This operator acts on the gravitational field as a \emph{displacement operator} whose amplitude is proportional to the optical field's intensity.

Let us proceed with the selection of specific states within the Hilbert space, commencing with the detector component.
To a very good approximation, the electromagnetic field inside the cavity is a monochromatic wave at frequency $\omega$. From the quantum point of view, we can model it as a single mode coherent state $\ket{\alpha}$ for some complex number $\alpha$. With this hypothesis, we can easily trace out the detector component of the state $\ket{\Psi\left(t\right)}$, and we are left with
\begin{eqnarray}
  \mathcal{E}\left(t\right) = \sqrt{\frac{\omega}{V_c}}\left(\frac{\alpha \mel{\Psi_{g}\left(t\right)}{\hat{\mathcal{D}}\left[q\eta\left(t\right)\right]}{\Psi_{g}\left(t\right)} + c.c. }{\sqrt{2}}\right) \,,
\end{eqnarray}
where now the (quantum) GW wavefunction $\ket{\Psi_{g}}$ enters into play independently. 

\subsection{Vacuum state}
Now, let us consider the possibility of preparing GW states in specific quantum states, with the simplest one being the vacuum state.
The vacuum state for the generic mode $\boldsymbol{k}$ (that is, a state with no gravitons of energy $\boldsymbol{k}$) can be written as:
\begin{equation}
  \ket{\Psi_{g}}=\ket{0_{\boldsymbol{k}}\left(t\right)} = e^{-i \hat{b}^\dagger_{\boldsymbol{k}} \hat{b}_{\boldsymbol{k}} \Omega_k t} \ket{0_{\boldsymbol{k}}} = \ket{0}\,,
\end{equation}
where we are following the standard harmonic oscillator convention of writing the eigenstates of the number operator $a_{\boldsymbol{k}}^\dagger a_{\boldsymbol{k}}$ as $\ket{n_{\boldsymbol{k}}}$, and $\ket{0}$ is the state with no gravitons at all. Therefore, to obtain the mean field we must evaluate the following expression:
\begin{eqnarray}
  \mathcal{E}\left(t\right) = \sqrt{\frac{\omega}{V}}\left(\frac{\alpha \prod_{\boldsymbol{k}} \mel{0}{\hat{\mathcal{D}}\left[q_{\boldsymbol{k}}\eta\left(\Omega_k t\right)\right]}{0} + c.c. }{\sqrt{2}}\right)\,.
\end{eqnarray}
The matrix element is easily calculated by considering that the vacuum can be seen as the coherent state with $\alpha=0$. The displacement operator acting on such a state gives
\begin{equation}\label{eq:displacementOperatorOnVacuum}
  \hat{\mathcal{D}}\left[q_{\boldsymbol{k}}\eta\left(\Omega_k t\right)\right] \ket{0} = \ket{q_{\boldsymbol{k}}\eta\left(\Omega_k t\right)}\,.
\end{equation}
Using the normalization condition for coherent states,\footnote{ 
The normalization condition is that given two coherent states characterized by complex numbers $\alpha$ and $\beta$, one has
\begin{equation}
  \braket{\alpha}{\beta} = e^{-\frac{1}{2}\left(\left|\alpha\right|^2+\left|\beta\right|^2 - \alpha^*\beta - \alpha\beta^* \right)}\,.
\end{equation}}
we can rewrite the previous expression as
\begin{eqnarray}
  \mel{0}{\hat{\mathcal{D}}\left[q_{\boldsymbol{k}}\eta\left(\Omega_k t\right)\right]}{0} = \braket{0}{q_{\boldsymbol{k}}\eta\left(\Omega_k t\right)} 
  = e^{-\frac{1}{2}q^2_{\boldsymbol{k}}\left|\eta\left(\Omega_k t\right)\right|^2}\,,
\end{eqnarray}
and thus define
\begin{equation}\label{eq:vacuumMatrixElement}
  D \defequal \prod_{\boldsymbol{k}}\mel{0}{\hat{\mathcal{D}}\left[q_{\boldsymbol{k}}\eta\left(\Omega_k t\right)\right]}{0} = e^{-\frac{1}{2}\sum_{\boldsymbol{k}}q^2_{\boldsymbol{k}}\left|\eta\left(\Omega_k t\right)\right|^2}\,.
\end{equation}
This quantity~\eqref{eq:vacuumMatrixElement} was calculated in~\cite{Guerreiro2020} by introducing both an infrared and an ultraviolet cutoff to avoid divergence and by noting that by simple algebra
\begin{equation}
  \left|\eta\left(\Omega_k t\right)\right|^2 = 2\left[1-\cos\left(\Omega_k t\right)\right] \,.
\end{equation}
The final result for the the mean field, in the case of a vacuum state, is
\begin{equation}\label{eq:fieldForVacuum}
  \mathcal{E}\left(t\right) = \sqrt{\frac{2\omega}{V_c}}D\Re{\alpha}\,.
\end{equation}
We can express this as equation~\eqref{eq"effectOnNoise}, which represents the effect of the detector's noise. Essentially, we have determined how the gravitational vacuum affects the interferometer's sensitivity curve. However, it's important to note that this impact is orders of magnitude below any reasonably achievable sensitivity limit \cite{Parikh2021a, Guerreiro2020}. In fact, it's even lower than the theoretical quantum thermal noise of gravity across a wide range of frequencies, as demonstrated in~\cite{Coradeschi2021}. Ultimately, this effect is impractical to measure.

When analyzed more closely, however, we do find an interesting side result. We performed this calculation after we had neglected the subdominant $q^2$ term in equation~\eqref{eq:evolutionOperatorFinal}. Had we retained it, we would have arrived to the surprising conclusion that the gravitational vacuum induces squeezing of the cavity field~\cite{Guerreiro2020}. Once again, after plugging in the right numbers, this effect turns out to be practically unmeasurable. 
However, in general, we find that, unsurprisingly, to achieve a measurable effect in an experiment where gravity couples to optical observables, one needs to start from gravity modes populated with a large mean number of gravitons, as we will see in the following.

\subsection{Coherent state}
The simplest state in which we can collect a large mean number of gravitons together is a coherent state. A calculation similar to the one we performed to arrive at equation~\eqref{eq:fieldForVacuum} can be carried out to derive the effect on the cavity's field quadrature of a single mode coherent state of gravity. We write such state as
\begin{equation}\label{eq:coherentStateGW}
  \ket{\Psi_{g}} = 
  \begin{cases}
    \ket{he^{i\Omega_{GW}t}} & \text{if } k=k_{GW}\,, \\
    \ket{0} & \text{otherwise}\,.
  \end{cases}
\end{equation}
Here $h$ is real (we set the phase to zero, for simplicity) and is indeed a large number linked to the population of the mode. Now we must calculate
\begin{equation}    
  \mathcal{E}\left(t\right) = \sqrt{\frac{\omega}{V_c}}\left(\frac{\alpha \mel{he^{i\Omega_{GW}t}}{\hat{\mathcal{D}}\left[q_{GW}\eta\left(\Omega_{GW} t\right)\right]}{he^{i\Omega_{GW}t}}\prod_{\boldsymbol{k}\neq \boldsymbol{k}_{GW}} \mel{0}{\hat{\mathcal{D}}\left[q_{\boldsymbol{k}}\eta\left(\Omega_k t\right)\right]}{0} + c.c. }{\sqrt{2}}\right)\,.
\end{equation}
When evaluating the $\boldsymbol{k}\neq \boldsymbol{k}_{GW}$ product, we would obtain a product of terms in $\sim e^{\frac{1}{2}q^2_{\boldsymbol{k}}}$ which - because of the small value of $q_{\boldsymbol{k}}$ - are all of order 1, much in the same way as we calculated expression~\eqref{eq:vacuumMatrixElement}. This means we can neglect all but
\begin{eqnarray}
  \mathcal{E}\left(t\right) \sim \sqrt{\frac{\omega}{V_c}}\left(\frac{\alpha \mel{he^{i\Omega_{GW}t}}{\hat{\mathcal{D}}\left[q_{GW}\eta\left(\Omega_{GW} t\right)\right]}{he^{i\Omega_{GW}t}} + c.c. }{\sqrt{2}}\right)\,.
\end{eqnarray}
Again, the full calculation is straightforward, arriving at
\begin{equation}
  \mathcal{E}\left(t\right) \sim e^{i2qh\sin\Omega t}\,.
\end{equation}
This expression is of the form~\eqref{eq:effectOnSignal}, and it tells us that we are measuring a signal oscillating in phase with the GW. We have thus recovered the classical GW signal from a representation of the quantum analog of a classical monochromatic wave: the single-mode coherent state. This gives us a first validation check of our program.

\subsection{GW-induced decoherence}\label{sec:decoherence}
In Sec.~\ref{sec:reconstruction}, we initially neglected the $q^2$ terms in the time evolution operator~\eqref{eq:evolutionOperatorFinal}. This choice was justified as these terms give rise to effects, such as the aforementioned squeezing of the cavity field induced by the gravitational quantum vacuum, that is by far dominated by effects that are linear in $q$.
However, it is worth paying some more attention to such $q$-quadratic terms, as it turns out they produce gravity-induced decoherence. 
Although the effect is weak and difficult to measure, it serves as a secondary validation check by linking our setup to established results.

To see this, let us now repeat the calculations presented in Sec.~\ref{sec:reconstruction} but this time by preparing our system as an electromagnetic (EM) qubit interacting with the gravitational vacuum
\begin{equation}\label{eq:EMQubitTimesGWVacuum}
  \ket{\Psi\left(0\right)} = \frac{\ket{0}_\text{EM}+\ket{N}_\text{EM}}{\sqrt{2}} \otimes \ket{0}_\text{GW}
\end{equation}
where, once again, $\ket{N}_\text{EM}$ ($\ket{N}_\text{GW}$) denotes a state with $N$ photons (gravitons). 
We can now perform the following three steps:
\begin{enumerate}
\item Evolve the state using the simplified time evolution operator as expressed in equation~\eqref{eq:displacementOperator};
\item write the total density matrix $\rho\left(t\right)$ associated to the time-evolving state;
\item calculate the density matrix $\rho_{EM}\left(t\right)$ associated to the EM subsystem, by tracing out the GW degrees of freedom.
\end{enumerate}
When carrying out the calculations, we arrive at the following denity matrix~\cite{Guerreiro2021}:
\begin{equation}\label{eq:densityMatrixTracedOut}
  \rho_\text{EM}\left(t\right) = \begin{pmatrix}
    \frac{1}{2} & \rho_{01} \\
    \rho_{01}^{*} & \frac{1}{2}
  \end{pmatrix}\,,
\end{equation}
where 
\begin{equation}
  \rho_{01}=\braket{0}{qN\eta} = e^{-\frac{1}{2}q^2N^2\left|\eta\right|^2}\,. 
\end{equation}
The presence of such time dependent off-diagonal terms (via the time dependence of $\eta\left(t\right)$) shows indeed that the $q^2$ component of the time evolution operator is inducing decoherence.

This can be taken farther. By replacing the vacuum $\ket{0}_\text{GW}$ in equation~\eqref{eq:EMQubitTimesGWVacuum} with a GW single mode coherent state $\ket{\alpha}_\text{GW}$, and repeating the same calculations, we obtain an EM density matrix of the same form as~\eqref{eq:densityMatrixTracedOut}, but now with 
\begin{equation}
  \rho_{01} = e^{-\frac{1}{2}\left[q^2N^2\left|\eta\right|^2+qN\left(\eta^*\alpha-\eta\alpha^*\right)\right]}\,.
\end{equation}
When we extend it to a single mode GWs in a thermal state, we obtain 
\begin{equation}\label{eq:rhoThermal}
  \rho_{01} = e^{-\frac{1}{2}q^2N^2\left|\eta\right|^2 \left(1+\overline{n}\right)}
\end{equation}
where $\overline{n}$ is the mean number of gravitons\footnote{
To arrive at equation~\eqref{eq:rhoThermal}, one should remember that the density matrix of a thermal state with mean number of gravitons $\overline{n}$, can be related to the continuum of coherent states $\ket{\alpha}$ as
\begin{equation}
  \rho = \int \frac{d^2\alpha}{\pi \overline{n}} e^{-\frac{\left|\alpha\right|^2}{\overline{n}}} \ket{\alpha}\bra{\alpha}\,.
\end{equation}
}. This result is useful in that it finally allows us to consider an ensemble of modes in thermal states. In such a case, we would need to reintroduce the state index $\boldsymbol{k}$, and calculate
\begin{eqnarray}
  &\ & \prod_{\bm{k}}  e^{-\frac{1}{2}q_{\bm{k}}^2N^2\left|\eta{\bm{k}}\right|^2\left(1+\overline{n}_{\bm{k}}\right)} = e^ {-\frac{1}{2}N^2 \sum_{\bm{k}} q_{\bm{k}}^2\left|\eta{\bm{k}}\right|^2\left(1+\overline{n}_{\bm{k}}\right)    }
\end{eqnarray}
For large temperatures $T$, one approximates $(1+\overline{n}_{\boldsymbol{k}})$ with $\overline{n}_{\boldsymbol{k}}$, and $ T $ simply counts the number of gravitons in each mode as
\begin{equation}
  \overline{n}_{\boldsymbol{k}} = \frac{k_BT}{\Omega_{\boldsymbol{k}}}\,.
\end{equation}
where $k_B$ is Boltzmann's constant. Using the explicit forms of $q_{\boldsymbol{k}}$ and $\eta_{\boldsymbol{k}}$, and averaging over the Bose-Einstein distribution (see~\cite{Guerreiro2021}), we arrive at an expression of the form
\begin{equation}
  \rho_{01} \approx e^{- \Gamma t}\,,
\end{equation}
where 
\begin{equation}
  \Gamma \propto k_{B}T \left( \dfrac{\Delta E}{E_{\rm pl}}\right)^{2}\,,
\end{equation}
and $\Delta E = N\omega$ is the energy of the state $\ket{N}_{\text{EM}}$, in accordance with previous results on gravitational-induced decoherence~\cite{Blencowe2013}.

\section{GW-induced electric field fluctuations}\label{sec:observables}
Before we continue discussing other gravity wave states, it is important to comment on the practicality of measuring deviations from the classical theory with elecrtomagnetic probes.
In a previous work~\cite{Guerreiro2021}, we showed that the measurement problem for our quantum gravitational states can be stated in terms of photon-number tomography of the optical mode that interacts with the wave. In particular, we showed that if the GW state is Gaussian, it can be reconstructed from experimentally accessible data that can be measured from non-classical (yet macroscopic) observables.
Here, we would like to point out that information on the GW states can also be obtained from field (homodyne) measurements, which is more practical than a photon-number resolving measurement experiments. 

Reconstruction of the second moments of a general GW state $ \vert \Psi \rangle $ can be achieved by measuring expectation values of the form $ \langle \Psi \vert \mathcal{D}(nq\eta(t)) \vert \Psi \rangle $, where $n$ is an integer \cite{Guerreiro2021}. General reconstruction of the first and second moments can be achieved if we measure these expectation values for $ n = 1, 2, 3$. For coherent states, this can be done by measuring the first three moments of the electric field.

The variance of the field is
$
\Delta \mathcal{E} = \langle \mathcal{E}^2 \rangle - \langle \mathcal{E} \rangle ^2
$
and we compute $\langle \mathcal{E}^2 \rangle$ by tracing out the detector component. Noticing that 
\begin{equation}
  \mathcal{E}^2(t) = \frac{\omega}{2V_c}\left(2a^\dagger a - 1+ a^2\mathcal{D}(-2q\eta(t)) + \left(a^\dagger\right)^2\mathcal{D}^*(-2q\eta(t)\right) ,
\end{equation}
we find that for general states, 
\begin{equation}
  \langle \mathcal{E}^2(t) \rangle =\frac{\omega}{2V_c} \left(|\alpha|^2 - 1 + \alpha^2\mel{\Psi_g(t)}{\hat{\mathcal{D}}(2q\eta(t)}{\Psi_g(t)} + c.c.\right)\,.
\end{equation}
Assuming that the gravitational wave is initially  a vacuum state, we have
\begin{equation}
  \label{eq:mean_square_EM}
  \langle \mathcal{E}^2\rangle = \frac{\omega}{2V_c}\left(\alpha^2\mel{0}{\hat{\mathcal{D}}\left[2q_{\boldsymbol{k}}\eta\left(\Omega_k t\right)\right]}{0}+{\alpha^*}^2\mel{0}{\hat{\mathcal{D}}^*\left[2q_{\boldsymbol{k}}\eta\left(\Omega_k t\right)\right]}{0}-1+|\alpha|^2\right)\,.
\end{equation}
Analogously, 
\begin{equation}
  \mel{0}{\hat{\mathcal{D}}\left[2q_{\boldsymbol{k}}\eta\left(\Omega_k t\right)\right]}{0} = \braket{0}{2q_{\boldsymbol{k}}\eta\left(\Omega_k t\right)}
  = e^{-{2}q^2_{\boldsymbol{k}}\left|\eta\left(\Omega_k t\right)\right|^2}\,,
\end{equation}
and we can define the quantity,
\begin{equation}\label{eq:vacuumMatrixElement2}
  D_2 \defequal \prod_{\boldsymbol{k}}\mel{0}{\hat{\mathcal{D}}\left[2q_{\boldsymbol{k}}\eta\left(\Omega_k t\right)\right]}{0} = e^{-2\sum_{\boldsymbol{k}}q^2_{\boldsymbol{k}}\left|\eta\left(\Omega_k t\right)\right|^2}\,.
\end{equation}
Using the previous equations, the mean square value of the electric field interacting with the GW~\eqref{eq:mean_square_EM} becomes
\begin{equation}
  \langle \mathcal{E}^2 \rangle = \frac{\omega}{2V_c}\left[ 2D_2\Re(\alpha^2) - 1+|\alpha|^2\right]\,.
\end{equation}
As was shown in \cite{Guerreiro2021}, to determine the second order correlation functions of the GW, we need to evaluate terms proportional up to $\mel{0}{\hat{\mathcal{D}}\left[3q_{\boldsymbol{k}}\eta\left(\Omega_k t\right)\right]}{0}$. In order to achieve that, we need to go up to the third order moment, the skewness ($\Delta s$), defined as
\begin{equation}
  \Delta s = \langle\mathcal{E}^3 \rangle - \langle \mathcal{E}\rangle \langle \mathcal{E}^2\rangle + 2\langle\mathcal{E}\rangle^3\,.
\end{equation}
Note that all the terms in the above definition have already been computed, except for $\langle \mathcal{E}^3\rangle$. We now turn our attention to this particular term.
Notice that
\begin{eqnarray}
  \mathcal{E}^3(t) &=& \left(\frac{\omega}{2V_c}\right)^{3/2}\left\{a^3\mathcal{D}[-3q\eta(t)] +(2a^\dagger a a - 3a)\mathcal{D}[-q\eta(t)] \right. \nonumber \\ 
  & & \left. + (2a^\dagger a^\dagger a - 3a^\dagger)\mathcal{D}^*[-q\eta(t)] + (a^\dagger)^3\mathcal{D}^*[-3q\eta(t)]\right\}\,.
\end{eqnarray}
For the initial vacuum state we find
\begin{equation}
  \langle \mathcal{E}^3(t)  \rangle = \left(\frac{\omega}{2V_c}\right)^{3/2}\left[\alpha^3\mel{0}{\hat{\mathcal{D}}\left[3q_{\boldsymbol{k}}\eta\left(\Omega_k t\right)\right]}{0}+\left(|\alpha|^2\alpha - 3\alpha\right)\mel{0}{\hat{\mathcal{D}}\left[q_{\boldsymbol{k}}\eta\left(\Omega_k t\right)\right]}{0} + c.c\right],
\end{equation}
and
\begin{equation}
  \langle \mathcal{E}^3  \rangle =\left (\frac{\omega}{2V_c}\right)^{3/2}\left[ 2D_3\Re(\alpha^3) - 2D\Re(\alpha)(3-|\alpha|^2)\right]\,,
\end{equation}
where $D_3$ is defined as
\begin{equation}
  D_3 \defequal \prod_{\boldsymbol{k}}\mel{0}{\hat{\mathcal{D}}\left[3q_{\boldsymbol{k}}\eta\left(\Omega_k t\right)\right]}{0} = e^{-\frac{9}{2}\sum_{\boldsymbol{k}}q^2_{\boldsymbol{k}}\left|\eta\left(\Omega_k t\right)\right|^2}\,.
\end{equation}
With this, we see that the quantities $ D, D_2 $ and $ D_3$ can be obtained from measurements of the first three moments of the electric field, which in turn can be measured via homodyne detection. In possession of these quantities, we can then reconstruct the first and second moments of GW vacuum fluctuations. This calculation can easily be extended to the case of a coherent sate. 
GW-induced electric field fluctuations can also be calculated for other states following the recipe introduced above, although in general perfect state reconstruction cannot be achieved from measurements of the electric field moments alone. These fluctuations could, however, lead to interesting signatures \cite{arani2023sensing}.

\section{Squeezed gravity}\label{sec:squeezing}
So far, we have considered the vacuum, coherent states, and thermal states for the gravity modes. While interesting theoretical insight can be gained by studying these cases, none of them will yield measurable quantum effect under our assumptions. It is still interesting (and, indeed, a required sanity check of our approach) that in the case of a highly populated coherent state, we recover the classical signal even though we started from a completely quantum picture.

The natural next step in our investigation is asking the question: can any quantum states of gravity exist that have no classical analog and have a chance of yielding a detectable effect? 

We give a tentative answer to this question by drawing from quantum optics experience and investigating what happens if we \emph{assume} that gravity can live in a \emph{squeezed state} -- the analogue of squeezed states of light that are routinely produced in optics laboratories. Before going on, we stress that this assumption implies the existence of some mechanism that does put gravity into such a state, which is as yet unknown for GWs in the LIGO band. To be more precise, there is at least one candidate in the category: an established consensus exists on the hypothesis that inflation might indeed have squeezed gravity at primordial times; however, once again -- much like in the cases of vacuum corrections or gravitational decoherence -- this effect leads to a weak signal~\cite{Coradeschi2021}. Nonetheless, calculating the potential signature of a gravitational squeezed vacuum on our model detector is instructive. After all, we cannot a priori exclude mechanisms, other than inflation, that could produce squeezing (see section~\ref{sec:sources} for more detailed discussion on this point).

Let us thus prepare gravity in a squeezed state, which we may model as a mode of the form
\begin{equation}
  \ket{\Psi_{g}} = \ket{\beta e^{2i\Omega t}} = \hat{S}\left(\beta\right)\ket{0}\,,
\end{equation}
where the complex number $\beta$ is the squeezing parameter and $\hat{S}\left(\beta\right)$ is the squeezing operator, as defined in textbooks. Now the matrix element for the mean electric field~\eqref{eq:electriFieldQuadratureOperatoreGWTraced} contains terms of the form
\begin{equation}
  \alpha \mel{0}{\hat{S}^\dagger\left(\beta\right)  \hat{\mathcal{D}}\left[q\eta\left(t\right)\right]\hat{S}\left(\beta\right)}{0}\,,
\end{equation}
and 
after some calculations, which were slightly more involved but nonetheless straightforward, we arrive at our result:
\begin{equation}
  \mathcal{E}\left(t\right) \sim 2\alpha \left[ 1-8q^2 e^{2\left|\beta\right|} \sin^4 \left(\frac{\Omega t}{2}\right)\right]\,.
\end{equation}
This term is of the form of equation~\eqref{eq"effectOnNoise} and it thus tells us that we found an effect on the \emph{noise}: a squeezed gravitational vacuum would manifest itself at an interferometer as an additional, oscillating term, in the noise spectrum of the instrument~\cite{Parikh2021a, Parikh2021b, Guerreiro2020}. The most interesting part is that the amplitude of such an oscillating term contains an exponential factor. Such a factor could behave as an enhancement term to the noise, depending on the magnitude of the squeezing parameter $\beta$. The magnitude of $ \beta$, in turn, depends on the details of the source dynamics, which at this moment we cannot foresee. Nonetheless, if mechanisms exist in nature that would produce such an exponentially enhanced effect, we cannot exclude that future, more sensitive detectors could actually see it. In the subsequent section, we shall delve deeper into this topic.

Something even more interesting happens when we prepare gravity in a \emph{squeezed-coherent} state. For a single mode, this would mean
\begin{equation}
  \ket{\Psi_{g}} = \hat{S}\left(\beta\right)\hat{\mathcal{D}}\left(he^{i\Omega_{GW}t}\right)\ket{0}\,,
\end{equation}
where we have used the same notation as in eq.~\eqref{eq:coherentStateGW}. Such a state would represent a squeezed quantum gravitational wave mode propagating from the source to the detector. The electromagnetic analog would be a squeezed laser beam, which we are able to generate and propagate in a laboratory.

After calculating the electric field matrix element as usual, for the first time we find an effect of the type described by equation~\eqref{eq:effectOnSignal} which deviates from the classical behavior. Specifically, after rewriting $\beta=re^{i\xi}$, we get
\begin{equation}~\label{eq:squeezedCoherentEffect}
  \delta\phi = 2 h q \left[ \sin\left(\Omega t\right) \cosh\left(r\right) + \sin\left(2\xi-\Omega t\right) \sinh\left(r\right) - \sin\left(2\xi\right) \sinh\left(r\right) \right]
\end{equation}
Thus, an exponentially enhanced or suppressed effect, but this time on the \emph{signal}. Once again, the magnitude of the effect depends on the dynamics at the source, which at this point remains unknown. Nonetheless, equation~\eqref{eq:squeezedCoherentEffect} clearly shows how a purely quantum effect (squeezing) involving a state with a macroscopically high number of gravitons $\left|h\right|$, would produce a signal which can be exponentially enhanced -- even to order one -- and can thus be detectable with current or near future technology~\cite{Guerreiro2021}.

\subsection{Are squeezed gravitational waves produced in nature?}\label{sec:sources}
In the framework in which we are working, where GR is seen as a classical limit of an intermediate-energy effective quantum field theory of (self)-interacting gravitons, squeezed gravitational waves definitely exist \emph{theoretically}, that is, they are allowed states in the Hilbert space of the theory.

The question however remains open on whether there are any realistic astrophysical sources that could produce GWs with a sizable (i.e. potentially measurable) amount of squeezing. While the aforementioned hypothesis on the squeezing of the relic gravitational background induced by inflation seems to be widely accepted, it is predicted to be too small to be observed at gravitational interferometers. Drawing on our experience on quantum optics, we can outline two basic conditions that we can expect to be met in order to have measurable squeezing in a physical process:
\begin{enumerate}
\item The process should involve states characterized by a high -- macroscopic -- occupancy number;
\item The resulting state should be capable of propagating (ideally) undisturbed from source to detector.
\end{enumerate}
Both conditions are naturally met by the processes that produce the gravitational waves that we are able to observe. First of all, there is no doubt that mergers of black holes (or really any other sources of high-intensity GWs) involve macroscopic number of gravitons (assuming, of course, that gravitons do exist). Furthermore, gravity is naturally weak-interacting at low energy, which means a GW basically stops interacting as soon as it leaves its source, traveling the distance to the detector nearly undisturbed. Note that this contrasts sharply with the behaviour of squeezed sources in electromagnetism: in quantum optics laboratories, squeezed beams of light are commonly produced by using intense laser beams, which are prone to losing coherence because of the high probability of interaction with any medium present in the laboratory. Transporting the quantum state of a laser beam (e.g. a squeezed coherent state) over long distances from source to detector is thus challenging, and can only be achieved with great effort in a laboratory. Gravitational waves can be expected to be free of this problem.

However, even if merger events (or other sources of strong GWs) do satisfy the minimum requirements to be candidate producers of squeezed gravitational waves, it is a challenge to understand if they \emph{actually} produce such states. Answering this question appears to involve the theoretical treatment of quantum effects at strong GR regimes, which is beyond our current capacities. However, while we are not (yet) capable of providing a formal argument in favour of the production of squeezed GWs at mergers and similar events (but for some discussion on the topic, see~\cite{Guerreiro2021}), we would still like to argue, more heuristically, that this possibility does deserve further investigation. Let us think once again of the analogy with quantum optics. Squeezed states of light are produced in the lab by making intense laser beams interact with anisotropic crystals. In fact, the mechanism that turns a coherent laser beam into a squeezed coherent laser beam involves the interaction of intense light with a highly non-linear optical medium. As electromagnetism is a linear theory at the classical level (nonlinear effects are only manifest in the quantum regime), producing squeezed light is in a sense an ``exotic'' process, in that it needs well-controlled laboratory conditions and is not found spontaneously in nature. Compare this with the case of gravity. Contrary to Maxwell's equations, Einstein's equations are already nonlinear at the classical level, and nonlinear effects affect phenomena taking place in strong regimes~\cite{nonlinear1, nonlinear2}. This means that squeezing of macroscopic gravitational waves might well be a natural effect, provided the source is strong enough -- which is definitely true in the case of mergers. Strong nonlinear astronomical sources, perhaps those already known to emit GWs, seem therefore reasonable candidates to investigate. A recent step in this direction has been proposed in~\cite{Guerreiro:2023gdy} where nonlinear effects present in black hole's ringdown~\cite{nonlinear1, nonlinear2,Kehagias:2023ctr, London:2014cma, Ma:2022wpv,Lagos:2022otp} have been considered.



\section{Conclusions}
\label{sec:conclusions}
In this paper, we have shown explicitly how to make use of quantum optics in order to derive phenomenological results in quantum gravity in the weak gravity regime. We focused on the treatment of the problem from the point of view of the equations of motion and applied the result to a model GW interferometer interacting with a few possible quantum states of gravity. We have examined various quantum states ranging from the basic vacuum state to the coherent state, and ultimately concluded with an evaluation of squeezed states.
Among the ones we evaluated, \emph{squeezed-coherent} gravitational waves have proven to be the most promising candidates for providing potentially detectable quantum aspects of gravity. 
%
%
The findings of Sec.~\ref{sec:sources}, however basic, together with the results reported in Sec.~\ref{sec:squeezing}, show how a squeezed coherent GW could produce an effect on the \emph{signal} of a GW interferometer, and that such an effect has the potential of being of order 1. This indicates to us that further research on the topic - especially regarding the existence of possible sources - has promise and is worth pursuing.  

\bibliographystyle{unsrt}
\bibliography{main}

\end{document}